\documentclass[twocolumn, prl, superscriptaddress]{revtex4-1}
\usepackage{graphicx}
\usepackage{epsfig}
\usepackage{color}
\usepackage{amsmath}
\usepackage{ulem}
\begin{document}
\title{Universal relations for ultracold reactive molecules}
\author{Mingyuan He}
\thanks{They contribute equally to this work.}
\affiliation{Department of Physics and Astronomy, Purdue University, West Lafayette, IN, 47907, USA}
\affiliation{Shenzhen JL Computational Science and Applied Research Institute, Shenzhen, 518000, China}
\affiliation{Department of Physics, The Hong Kong University of Science and Technology, Clear Water Bay, Kowloon, Hong Kong, China}
\author{Chenwei Lv}
\thanks{They contribute equally to this work.}
\affiliation{Department of Physics and Astronomy, Purdue University, West Lafayette, IN, 47907, USA}
\thanks{They contribute equally to this work.}
\author{Hai-Qing Lin}
\affiliation{Beijing Computational Science Research Center, Beijing, 100193, China}
\author{Qi Zhou}
\email{zhou753@purdue.edu}
\affiliation{Department of Physics and Astronomy, Purdue University, West Lafayette, IN, 47907, USA}
\affiliation{Purdue Quantum Science and Engineering Institute, Purdue University, West Lafayette, IN, 47907, USA}
\date{\today}
\begin{abstract}
The realization of ultracold polar molecules in laboratories has pushed both physics and chemistry to new realms \cite{ Ye2, Deiglmayr, Cornish1, Nagerl2, DeMille4, Zwierlein1, Wang2, Doyle}. In particular, these polar molecules offer scientists unprecedented opportunities to explore chemical reactions in the ultracold regime where quantum effects become profound \cite{Ye3, Gregory, Wang, Ye1, Ni}. However, a key question about how two-body losses depend on quantum correlations in an interacting many-body system remains open so far. Here, we present a number of universal relations that directly connect two-body losses to other physical observables, including the momentum distribution and density correlation functions. These relations, which are valid for arbitrary microscopic parameters, such as the particle number, the temperature, and the interaction strength, unfold the critical role of contacts, a fundamental quantity of dilute quantum systems \cite{Tan2} in determining the reaction rate of quantum reactive molecules in a many-body environment. Our work opens the door to an unexplored area intertwining quantum chemistry, atomic, molecular and optical physics, and condensed matter physics.
\end{abstract}

\maketitle
 
In a temperature regime down to a few tens of nano-Kelvin, highly controllable polar molecules provide scientists with a powerful apparatus to study a vast range of new quantum phenomena in condensed matter physics, quantum information processing and quantum chemistry, such as exotic quantum phases \cite{Buchler1, Cooper1, Yao, Syzranov}, quantum gates with fast switching times \cite{DeMille1, Yelin}, and quantum chemical reactions \cite{Ye3, Gregory, Wang, Ye1, Ni}. In all these studies, the two-body loss is an essential ingredient leading to non-hermitian phenomena. Similar to other chemical reactions, collisions between molecules may yield certain products and release energies, which allow particles to escape the traps. For instance, a prototypical reaction, $\text{KRb+KRb}\rightarrow \text{K}_2+\text{Rb}_2$, is the major source causing the loss of KRb molecules. Undetectable complexes may also form, resulting in losses in the system of interest \cite{Wang, Gregory}.

Whereas chemical reactions are known for their complexities, taking into account quantum effects imposes an even bigger challenge to both physicists and chemists. The exponentially large degrees of freedom and quantum correlations built upon interactions make it difficult to quantitatively analyze the reactions. A standard approach is to consider two interacting particles, the reaction rate of which is trackable \cite{Mayle, Julienne1}. Though such results are applicable in many-body systems when the temperature is high enough and correlations between different pairs of particles are negligible, with decreasing the temperature, many-body correlations become profound and this approach fails. A theory fully incorporating quantum many-body effects is desired to understand the chemical reaction rate.

In this work, we show that universality exists in chemical reactions of ultracold reactive molecules. We implement contacts, the central quantity in dilute quantum systems \cite{Tan2}, to establish universal relations between the two-body loss rate and other quantities including the momentum distribution and the density correlation function. Previously, two-body losses of zero-range potentials hosting inelastic s-wave scatterings were correlated to the s-wave contact \cite{Eric1, Salomon}. In reality, chemical reactions happen in a finite range. Many systems are also characterized by high-partial wave scatterings. For instance, single-component fermionic KRb molecules interact with p-wave scatterings \cite{Ye1,Ye2}. It is thus required to formulate a theory applicable to generic short-range reactive interactions. To concretize discussions, we focus on single-component fermionic molecules. All our results can be straightforwardly generalized to other systems with arbitrary short-range interactions. 

\begin{figure}
\centering
\includegraphics[width=2.8in]{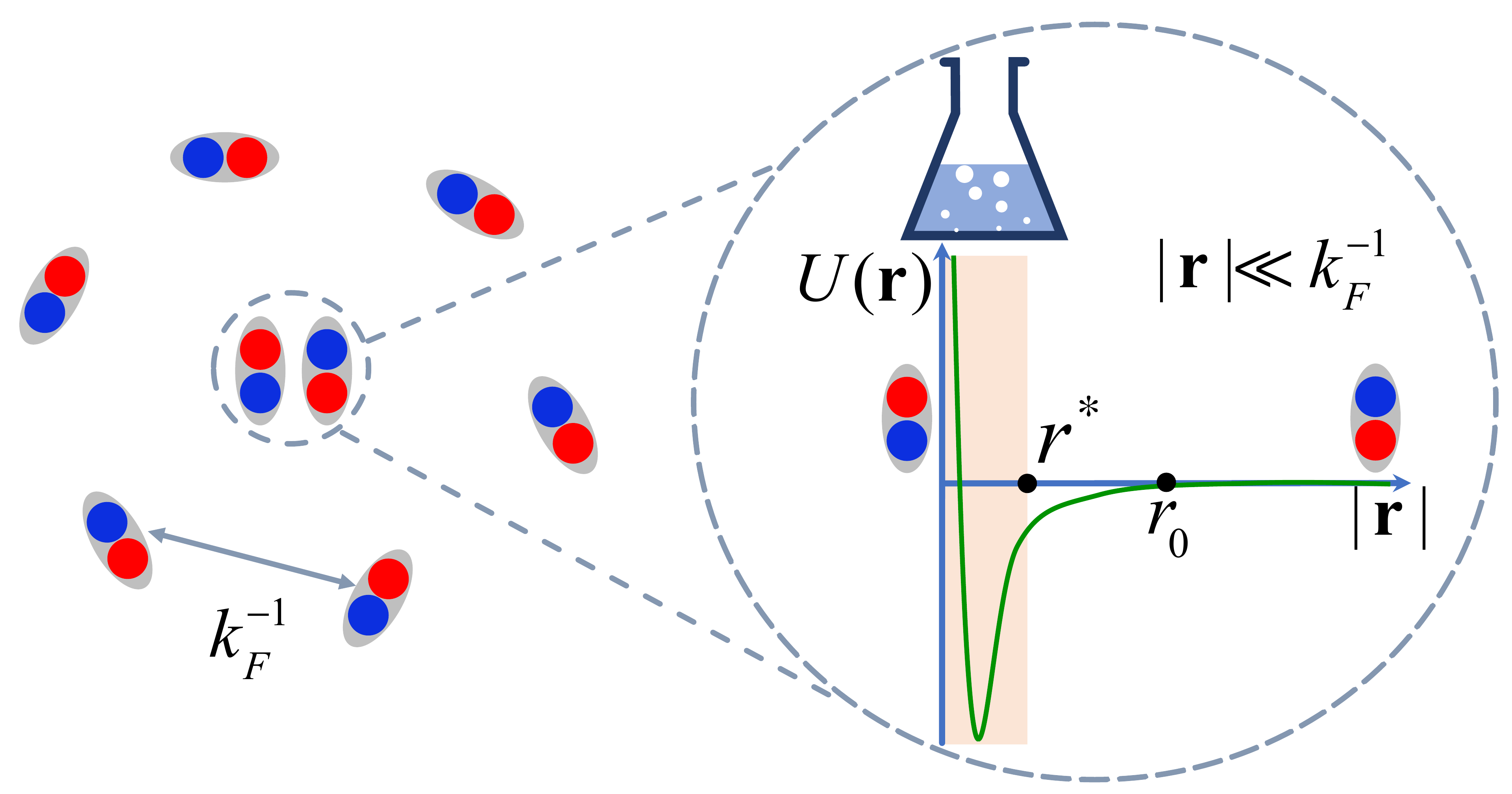}
\caption{A length scale separation in dilute molecules. The blue (red) solid spheres represent potassium (rubidium) atoms. Inside the dashed circle are two molecules, the separation between which is much smaller than the average inter-particle spacing, $|{\bf r}|\ll k_F^{-1}$. The enlarged plot of the regime inside the dashed circle is a schematic of the chemical reaction. The green solid curve represents the real part of the interaction, $U_R({\bf r})$. The imaginary part of the interaction, $U_I({\bf r})$, is nonzero only in the shaded area, where the reaction happens. }\label{Fig1}
\end{figure}

The Hamiltonian of $N$ reactive molecules is written as 
\begin{equation}
H=\sum_i [-\frac{\hbar^2}{2M}\nabla_i^2+V_{\rm ext}({\bf r}_i)]+\sum_{i> j} U({\bf r}_i-{\bf r}_j), \label{H}
\end{equation}
where $M$ is the molecular mass, $V_{\rm ext}({\bf r})$ is the external potential, $U({\bf r})$ is a two-body interaction, as shown in Fig. \ref{Fig1}. The many-body wavefucntion, $\Psi({\bf r}_1,{\bf r}_2, ..., {\bf r}_N )$, satisfies the time-dependent Schr\"odinger equation, 
\begin{equation}
i\hbar \partial_t \Psi({\bf r}_1,{\bf r}_2, ..., {\bf r}_N )=H\Psi({\bf r}_1,{\bf r}_2, ..., {\bf r}_N ).\label{wf}
\end{equation}
In the absence of electric fields, $U({\bf r})$ is a short-range interaction with a characteristic length scale, $r_0$. When $|{\bf r}|>r_0$, $U({\bf r})=0$. Chemical reactions happen in an even shorter length scale, $r^*<r_0$. We adopt the one-channel model using a complex $U({\bf r})=U_R({\bf r})+i U_I({\bf r})$ to describe the chemical reaction \cite{Julienne1}, where $U_I({\bf r})\leq0$. When $|{\bf r}|>r^*$, $U_I({\bf r})=0$. Using the Lindblad equation that models the losses by jump operators, the same universal relations can also be derived (Supplementary Materials). 

Universal relations arise from a length scale separation in dilute quantum systems, $r^*<r_0\ll k_F^{-1}$, where $k_F^{-1}$, the inverse of the Fermi momentum, captures the average inter-particle separation. When the distance between two molecules is much smaller than $k_F^{-1}$, we obtain,
\begin{equation}
\Psi({\bf r}_1,{\bf r}_2, ..., {\bf r}_N ) \stackrel{|{\bf r}_{ij}|\ll k_F^{-1}}{\xrightarrow{\hspace*{1cm}} }  \sum_{m,\epsilon}  \psi_{m}({\bf r}_{ij};\epsilon)G_{m}({\bf R}_{ij};E-\epsilon), \label{Asym}
\end{equation}
where ${\bf r}_{ij}={\bf r}_i-{\bf r}_j$ denotes the relative coordinates of the $i$th and the $j$th molecules, ${\bf R}_{ij}=\{({\bf r}_i+{\bf r}_j)/2,{\bf r}_{k\neq i,j}\}$ is a short-hand notation including coordinates of their center of mass and all other particles. $\psi_{m}({\bf r}_{ij};\epsilon)$ is a p-wave wavefunction with a magnetic quantum number $m=0, \pm 1$, which is determined solely by the two-body Hamiltonian, $H_2=-({\hbar^2}/{M})\nabla^2+U({{\bf r}_{ij}})$, as all other particles are far away from the chosen pair in the regime, $|{\bf r}_{ij}|\ll k_F^{-1}$. $E$ is the total energy of the many-body system. $\epsilon=\hbar^2q_\epsilon^2/M$, the colliding energy, is no longer a good quantum number in a many-body system, and a sum shows up in Eq. (\ref{Asym}). Since both a continuous spectrum and discrete bound states may exist, we use the notation of sum other than an integral.

Though $\psi_{m}({\bf r}_{ij};\epsilon)$ depends on the details of $U({\bf r}_{ij})$ when $|{\bf r}_{ij}|<r_0$, it is universal when $r_0<|{\bf r}_{ij}|\ll k_F^{-1}$, as a result of vanishing interaction in this regime. We define $\psi_{m}({\bf r}_{ij};\epsilon)=\varphi_{m}(|{\bf r}_{ij}|;\epsilon)Y_{1m}(\hat{\bf r}_{ij})$, where $Y_{1m}(\hat{\bf r}_{ij})$ is the p-wave spherical harmonics. Whereas many resonances exist, the phase shift of the scattering between KRb molecules is still a smooth function of the energy, due to the large average line width of these resonances, which far exceeds the mean level spacing of the bound states \cite{Mayle}. The phase shift, $\eta$, then has a well defined expansion, $q_\epsilon^3 \cot[\eta(q_\epsilon)]=-1/v_p+q_\epsilon^2/r_e$, where $v_p$ and $r_e$ are the p-wave scattering volume and effective range, respectively, both of which are complex for reactive interactions. Consequently, $\varphi_{m} (|{\bf r}_{ij}|;\epsilon)=\varphi_{m}^{(0)}(|{\bf r}_{ij}|)+q_\epsilon^2\varphi_{m}^{(1)}(|{\bf r}_{ij}|)+{\rm O}(q_\epsilon^4)$, where
\begin{eqnarray}
&\varphi_{m}^{(0)}(|{\bf r}_{ij}|)& \stackrel{r_0< |{\bf r}_{ij}|\ll k_F^{-1}}{\xrightarrow{\hspace*{1.5cm}} } \frac{1}{{|{{\bf r}_{ij}}{|^2}}} - \frac{1}{{{v_p}}}\frac{{|{{\bf r}_{ij}}|}}{3},\label{wf2b1}\\
&\varphi_{m}^{(1)}(|{\bf r}_{ij}|)& \stackrel{r_0< |{\bf r}_{ij}|\ll k_F^{-1}}{\xrightarrow{\hspace*{1.5cm}} } \frac{1}{r_{e}}\frac{{|{{\bf r}_{ij}}|}}{3} + \frac{1}{{{v_p}}}\frac{{|{{\bf r}_{ij}}{|^3}}}{{30}} + \frac{1}{2}\label{wf2b2}.
\end{eqnarray}
To simplify expressions, we have considered isotropic p-wave interactions, $\varphi(|{\bf r}_{ij}|)=\varphi_{m}(|{\bf r}_{ij}|)$ and $G({\bf R}_{ij};E-\epsilon)=G_{m}({\bf R}_{ij};E-\epsilon)$, and suppressed other partial waves in the expressions, which do not show up in universal relations relevant for single-component fermionic molecules.

Using Eqs. (\ref{H}, \ref{wf}, \ref{Asym}), we find that the decay of the total particle number is captured by,
\begin{equation}
\partial_t N= -\frac{\hbar}{8\pi^2 M}\sum_{\nu=1}^3 \kappa_\nu C_\nu, \label{decay1}
\end{equation}
where the three contacts are written as
\begin{eqnarray}
C_{1}&=& 3(4\pi)^2N(N-1)\int d{\bf R}_{ij} |g^0|^2,\label{contact1}\\
C_{2}&=& 6(4\pi)^2N(N-1)\int d{\bf R}_{ij} {\rm Re}(g^{0*}g^1),\label{contact2}\\
C_{3}&=& 6(4\pi)^2N(N-1)\int d{\bf R}_{ij} {\rm Im}(g^{0*}g^1), \label{contact3}
\end{eqnarray}
$\int d {\bf R}_{ij}= \int d[({{\bf r}_{i}+{\bf r}_j})/{2}]d{\bf r}_{k\neq i, j}$ and $g^s=\sum\nolimits_\epsilon {q_\epsilon^{2s}G({\bf R}_{ij};E-\epsilon)}$. As shown later, $C_1$ determines the leading term in the large momentum tail, similar to systems without losses \cite{Ueda2, Zhang1, Zhou2, Zhang2}. In contrast, $C_{2,3}$ are new quantities in systems with two-body losses.

\begin{figure*}
\centering
\includegraphics[width=6.9in]{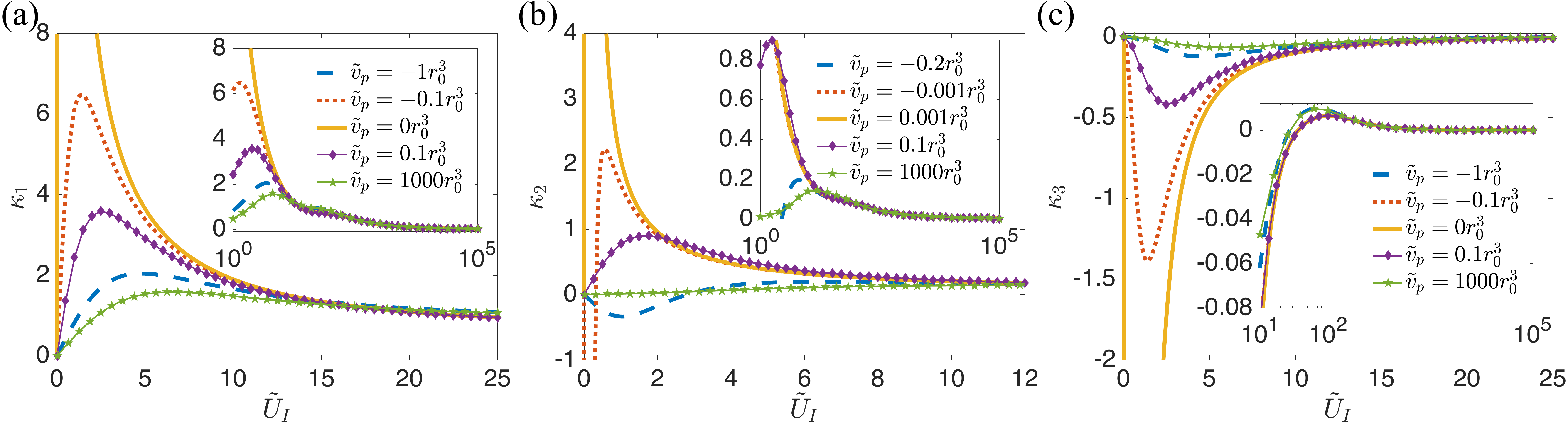}
\caption{Dependence of the three microscopic parameters on interactions. $\tilde v_p$ represents the scattering volume when $\tilde U_I=0$. $\tilde U_I$ is in the unit of $\hbar^2/(Mr_0^2)$. $\kappa_{1,2,3}$ are in unit of $r_0^{-3}$, $r_0^{-1}$ and $r_0^{-1}$, respectively. When $\tilde v_p$ crosses zero, the location of the maximum of $\kappa_1$ ($\kappa_3$) first approaches and then leaves the origin, and $\kappa_1$ ($\kappa_3$) remains positive (negative). In contrast, $\kappa_2$ quickly changes from positive to negative at small values of $\tilde U_I$ when $\tilde v_p$ crosses zero. In the large $\tilde U_I$ limit, all three parameters vanish, as shown by the insets. } \label{Fig2}
\end{figure*}

$\kappa_\nu$ in Eq. (\ref{decay1}) are microscopic parameters determined purely by the two-body physics. In our one-channel model, their explicit expressions are given by
\begin{eqnarray}
\kappa_1&=&-\frac{M}{\hbar^2}\int_0^\infty  {{U_I}\left( r \right){{\left| {\varphi ^{(0)}(r)} \right|}^2 r^2}dr} ,\\
\kappa_2&=&-\frac{M}{\hbar^2}{\rm{Re}} \left({\int_0^\infty  {{U_I}\left( r \right)\varphi ^{(0)*}(r)\varphi ^{(1)}(r) r^2dr} }\right), \\
\kappa_3&=&\frac{M}{\hbar^2}{\rm{Im}} \left({\int_0^\infty  {{U_I}\left( r \right)\varphi ^{(0)*}(r)\varphi ^{(1)}(r) r^2dr} }\right),
\end{eqnarray}
where $r=|{\bf r}|$. If $U({\bf r})$ is modeled by two square well potentials, one for its real part and the other for its imaginary part, $\kappa_{1,2,3}$ can be evaluated explicitly. For simplicity, we set $U({\bf r})=-\tilde U_R -i \tilde U_I$ when $|{\bf r}|\leq r_0=r^*$ and 0 elsewhere. Changing the ratio $r_0/r^*$ does not change any results qualitatively. Figure \ref{Fig2} shows how $\kappa_{1,2,3}$ depend on $\tilde U_I$ when $\tilde U_R$ is fixed at various values including those corresponding to small and divergent $v_p$ in the absence of $\tilde U_I$. When $\tilde U_I=0$, $\kappa_{1,2,3}=0$. With increasing $\tilde U_I$, $\kappa_{1,2,3}$ change non-monotonically and all approach zero when $\tilde U_I$ is large, indicating a vanishing reaction rate in the extremely large $U_I$ limit.

Equation (\ref{decay1}) is universal for any particle number and any short-range interactions with arbitrary interaction strengths, as well as any real external potential. It separates $C_\nu$, which fully capture the many-body physics, from two-body parameters, $\kappa_\nu$, which are independent on the particle number and the temperature. Therefore, even when microscopic details of the reactive interaction, for instance, the exact expression of $U({\bf r})$, are unknown, $\kappa_\nu$ can still be accessed in systems whose $C_\nu$ are easily measurable (Supplementary Materials). Equation (\ref{decay1}) also holds for any many-body eigenstates and a thermal average does not change its form. Therefore, Eq. (\ref{decay1}) does apply for any finite temperatures, provided that the reaction rate is slow compared to the time scale of establishing quasi-equilibrium in the many-body system, i.e.,  the many-body system has a well defined temperature at any time. Under this situation, $C_\nu$ should be understood as their thermal averages.

Interestingly, we have found that $\kappa_1$ and $\kappa_2$ can be rewritten as familiar parameters. In fact, $\kappa_1={\rm Im} {(v^{-1}_p)}$ and $\kappa_2={\rm Im} [-1/(2r_{e})]$ (Supplementary materials). In contrast, to our best knowledge, $\kappa_3$ is a new parameter that has not been addressed in previous works. Similar to $\kappa_2$, $\kappa_3$ can be expressed as the difference between the extrapolation of the two-body wavefunction in the regime $|{\bf r}|>r_0$ toward the origin and the realistic wavefunction at short distance, $|{\bf r}|<r_0$ (Supplementary materials). Equation (\ref{decay1}) can  be rewritten as
\begin{equation}
\partial_t N= -\frac{\hbar}{8\pi^2 M}\Big[\text{Im} ({v^{-1}_p})C_1- \frac{1}{2}\text{Im} (r_{e}^{-1}) C_2+\kappa_3 C_3\Big]. \label{decay2}
\end{equation}
For s-wave inelastic scatterings due to complex zero-range interactions, the first term on the right hand side of Eq. (\ref{decay2}) was previously derived, with $v^{-1}_p$ replaced by the complex s-wave scattering length \cite{Eric1}. For a generic short-range interaction, all three contacts and all three microscopic parameters are required, as shown in Eqs. (\ref{decay1}, \ref{decay2}). These expressions allow us to directly connect the two-body loss rate to a wide range of physical quantities.

We first consider the momentum distribution, which has a universal behavior when $|{\bf k}|\ll 1/r_0$ but much larger than all other momentum scales, including $k_F$, the inverses of the scattering length and the thermal wavelength. We define the total angular averaged momentum, ${n(|{\bf k}|)=\sum_{m=0,\pm 1}\int d{\bf\Omega} n_m({\bf k}) }$, where $\bf \Omega$ is the solid angle,
\begin{equation}
n(|{\bf k}|)\rightarrow  \frac{C_{1}}{|{\bf k}|^2}.
\label{nk}
\end{equation}
Once $n(|{\bf k}|)$ is measured, the first term in Eqs. (\ref{decay1}, \ref{decay2}) is known. If an $rf$ spectroscopy exists for molecules, similar to that for atoms, Eq. (\ref{nk}) also indicates that such spectroscopy has a universal tail, $\Gamma(\omega)\rightarrow [(\Omega_{rf}V)/(8\pi^2)]C_1 (\hbar\omega/M)^{-1/2}$, where $\omega$ is the $rf$ frequency, $\Omega_{rf}$ is the $rf$ Rabi frequency, and $V$ is the volume of the system. It is worth mentioning that, for atoms with elastic p-wave interactions, Eq. (\ref{nk}) describes the leading term of the large momentum tail \cite{Ueda2, Zhang1, Zhou2, Zhang2}. We have not found that the subleading term $\sim |{\bf k}|^{-4}$ has connections to two-body losses. 

Another fundamentally important quantity in condensed matter physics is the density correlation function, $S({\bf r})=\int d{\bf R} \langle n({\bf R}+{\bf r}/2)n({\bf R}-{\bf r}/2)\rangle$, which measures the probability of having two particles separated by a distance ${\bf r}$. Using Eqs. (\ref{Asym}, \ref{wf2b1}, \ref{wf2b2}), $S({\bf r})$ can be evaluated explicitly in the regime, $r_0<|{\bf r}|\ll k_F^{-1}$. To enhance the signal-noise-ratio, $S({\bf r})$ can be integrated over a shell with inner and outer radii, $x$ and $x+D$, respectively. Such an integrated density correlation is given by $P(x,D)= \int_{x}^{x+D} d{\bf r} S({\bf r})$, and 
\begin{equation}
    \begin{split}
        &\left.{\frac{\partial P(x, D)}{\partial D}}\right|_{D\to0} =\frac{1}{16\pi^2} \Big\{ C_1\frac1{x^2}+\frac{1}{2}C_2\\
        &\quad\quad-\Big[2{\rm Re}(\frac1{v_p})C_1-{\rm Re}(\frac1{r_e})C_2+{\rm Im}(\frac1{r_e})C_3\Big]\frac{x}{3} \Big\}.  \label{dcin}
    \end{split}
\end{equation}
Again, other partial-waves have been suppressed in the expression as their contributions are given by different spherical harmonics. Fitting $\left.{{\partial P(x, D)}/{\partial D}}\right|_{D\to0}$ measured in experiments using the power series in Eq. (\ref{dcin}) allows one to obtain all three contacts, $C_{1,2,3}$, provided that $v_p$ and $r_{e}$ are known. If these two parameters are unknown, it is necessary to include higher order terms in the expansion (Supplementary Materials). 

We emphasize that, no matter whether thermodynamic quantities and correlation functions can be computed accurately in theories, equations (\ref{decay1}, \ref{decay2}, \ref{nk}, \ref{dcin}) allow experimentalists to explore how contacts determine chemical reactions in interacting few-body and many-body systems. In fact, in the strongly interacting regime where exact theoretical results are not available, these universal relations become most powerful. 

\begin{figure}
  \centering
  \includegraphics[width=3.4in]{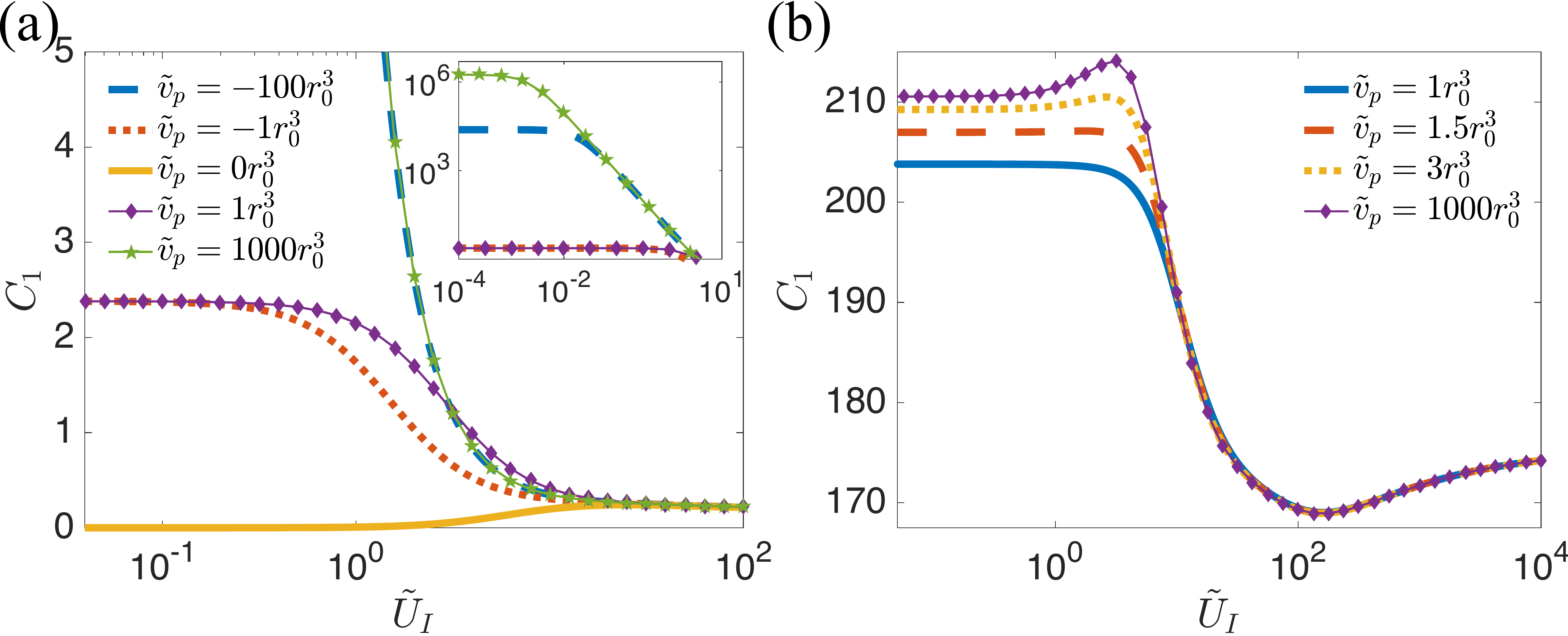}
  \caption{Contacts of a two-body system. (a) $C_1$ (in unit of $r_0^4/V$) of a scattering state as a function of $\tilde U_I$ (in unit of $\hbar^2/(Mr_0^2)$) when $\tilde v_p$ is fixed at various values. $q_\epsilon=0.01/r_0$. (b) $C_1$ (in unit of $r_0$) of a bound state as a function of $\tilde U_I$ (in unit of $\hbar^2/(Mr_0^2)$).  }\label{Fig3_1}
\end{figure}

It is useful to illuminate our results using some examples. For a two-body system in free space, the center of mass and the relative motion are decoupled. $\epsilon$ in Eqs. (\ref{contact1}, \ref{contact2}, \ref{contact3}) becomes a good quantum number, i.e., $G({\bf R}_{ij};E-\epsilon)$ becomes a delta function in the energy space. For scattering states with $\epsilon>0$, we consider the wavefunction, $\Psi^{[2]}({\bf r}_1,{\bf r}_2)=\phi_c({\bf R}_{12})\psi({\bf r}_{12})$, where $\phi_c({\bf R}_{12})$ is a normalized wavefunction of the center of mass and $\psi({\bf r}_{12})=\sqrt{8\pi/V}[{i}/({\cot{\eta}-i})][\cot{\eta} j_1 (q_\epsilon|{\bf r}_{12}|) -n_1(q_\epsilon|{\bf r}_{12}|)] \sum_m{Y_{1m}(\hat{\bf r}_{12})}$. Figure \ref{Fig3_1} shows the dependence of $C_1$ on $\tilde U_I$ when $\tilde v_p$ is fixed at various values. Results for a bound state are also shown. With increasing $\tilde U_I$, $C_1$ approaches a non-zero constant in both cases. Here, $C_2=2C_1 {\rm Re}(q_\epsilon^2)$. $C_3=0$ if we consider a scattering state. In contrast, $C_3=2C_1 {\rm Im}(q_\epsilon^2)$ for a bound state. Analytical results in the limits, $v_p= 0^{\pm}, \infty$, are shown in Table \ref{fig_table1}.

\begin{table} 
  \includegraphics[width=0.48\textwidth]{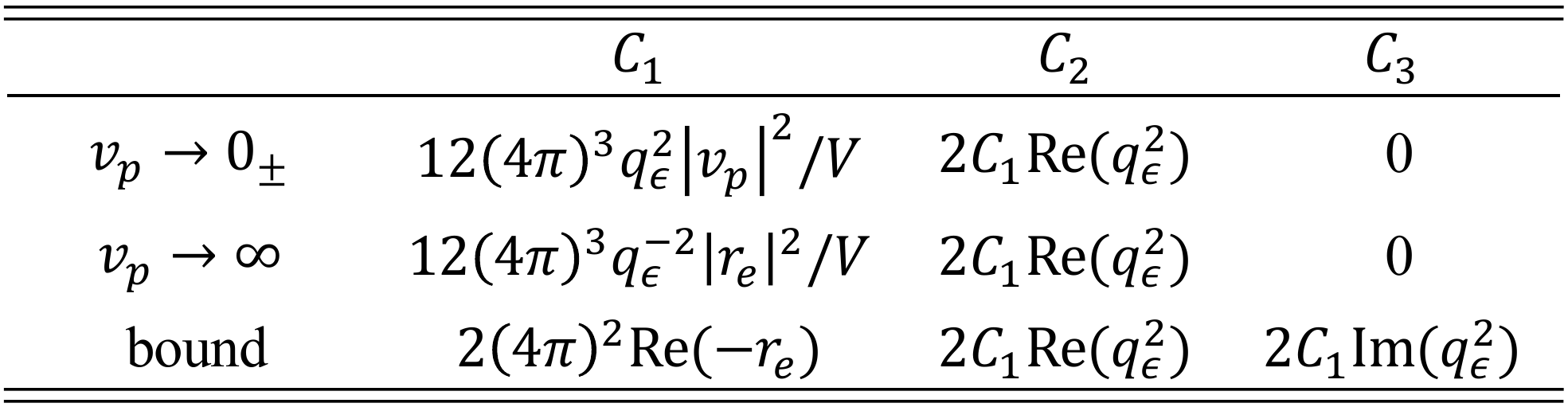}
  \caption{Analytical expressions for contacts $C_\nu$ of two particles in different limits. Line 1 and line 2 show the results in the weakly interacting regime and those at resonance, respectively. $v_p\rightarrow 0_\pm$ ($\infty$) means $v_p\rightarrow 0_\pm+0i$ ($\infty+0i$) on the complex plane. Line 3 includes the results for bound states, in which a single angular momentum $m$ is considered. }\label{fig_table1}
\end{table}

We use the second order virial expansion to study a thermal gas at high temperatures. The partition function is written as $Z=Z_0+e^{2\mu/(k_B T)}\sum_{E_c, n} (e^{-(E_c+\epsilon_n)/(k_B T)}- e^{-(E_c+\epsilon^0_n)/(k_B T)})$, where $Z_0$ is the partition function of non-interacting fermions, $\mu$ is the chemical potential, $E_c=\hbar^2 K^2/(4M)$ is the energy of the center of mass motion carrying a momentum $K$. $\epsilon_n$ and $\epsilon^0_n$ are the eigenenergies of the relative motion with and without interactions, respectively. Based on results of the two-body problem, thermal averaged contacts are derived using $\langle C_\nu\rangle_T=Z^{-1}e^{2\mu/(k_B T)}(\sum_{E_c}e^{-E_c/(k_B T)})(\sum_n C_\nu(\epsilon_n)e^{-\epsilon_n/(k_B T)})$. Using $N=k_B T\partial_\mu \ln Z$, we eliminate $\mu$ and obtain $\langle C_\nu\rangle_T$ as a function of $N$ and $T$. Analytical expressions in the limits, $v_p=0^{\pm}, \infty$, are shown in Table \ref{fig_table2}.  

\begin{table}[t]
  \includegraphics[width=0.48\textwidth]{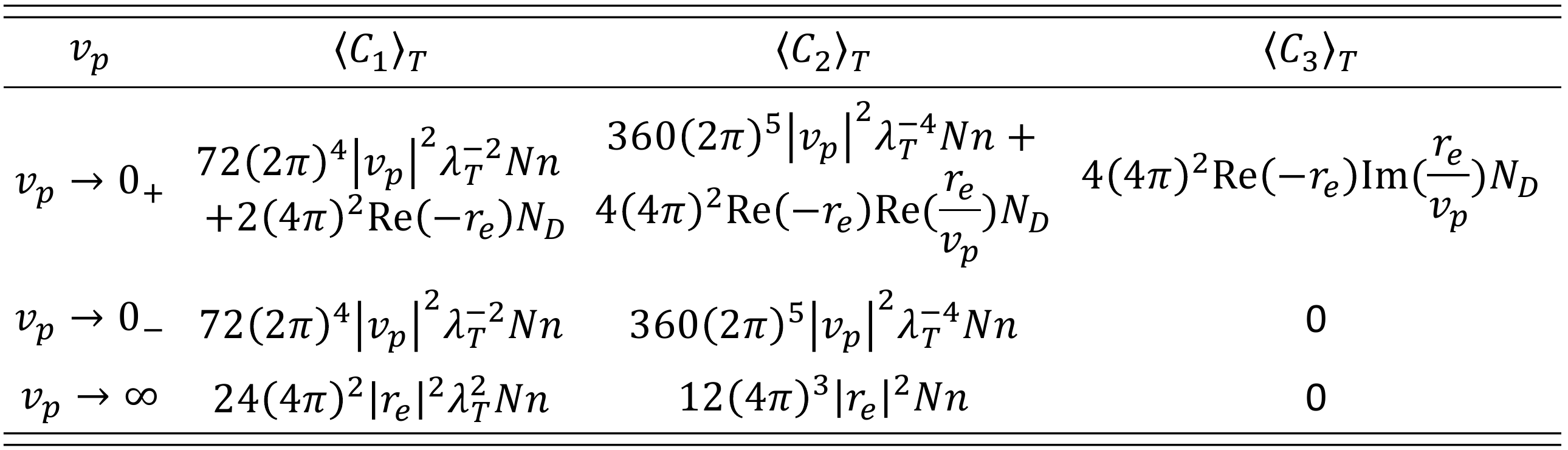}
  \caption{Analytical expressions for thermal averaged contacts $\langle C_\nu\rangle_T$ in different limits. Line 1 and 2 show the results in the weakly interacting regime. When $v_p$ is positive, bound states exist and their contributions  are included in Line 1. Line 3 includes the results at resonance. $N_D$ is the number of dimers. $\lambda_T = [(2\pi\hbar^2) /(M k_B T)]^{1/2}$ is the thermal wave length. }\label{fig_table2}
\end{table}

Table \ref{fig_table2} may shed light on some recent experiments conducted in the weakly interacting regime \cite{Ye1, Ye3}. Though $v_p>0$, it is likely that bound states are not occupied, i.e., the system is prepared at the upper branch. Therefore, $C_3=0$. In a homogenous system, we obtain, 
\begin{equation}
\begin{split}
\partial_t N=&\frac{144\pi^2}{h}{\rm Im}(v_p) Nn k_BT\\
&+\frac{360\pi^2}{h} {\rm Im}(\frac{v_p}{v_p^*}r_e^{-1}) \frac{M|v_p|^2}{\hbar^2} Nn k_B^2 T^2. \label{dh}
\end{split}
\end{equation}
A previous work derived the first term in Eq. (\ref{dh}) using a different approach \cite{Julienne1}. However, a complete expression needs to include the contribution from $r_e$, which leads to a different power of the dependence on $T$. A recent experiment has shown the deviation from the linear dependence on $T$ \cite{Ye1}. However, it is worth investigating whether such deviation comes from the second term in Eq. (\ref{dh}) or some other effects, in particular, correlations beyond the description of the second order virial expansion.

As a harmonic trap exists in experiments, the dependence on $T$ could be completely different. We use the local density approximation to obtain the total contacts by integrating local contacts. As a result, $C_\nu^{\rm trap}=[(\pi k_B T)/(M\omega^2)]^{3/2}{\cal C}_\nu(0)$, where $\omega$ is trapping frequency, and ${\cal C}_\nu(0)$ are the contact densities at the center of the trap (Supplementary Materials). Consequently,
\begin{equation}
  \begin{split}
  &\partial_t N^{\rm trap}=\frac{18\sqrt{\pi}}{h}(M{\omega}^2)^{\frac3{2}} {\rm Im}(v_p)(N^{\rm trap})^2 \frac{1}{\sqrt{k_BT}} \\
  &\quad+\frac{45\sqrt{\pi}}{h}  (M{\omega}^2)^{\frac3{2}}  {\rm Im} (\frac{v_p}{v_p^*} r_e^{-1})  \frac{M|v_p|^2}{\hbar^2}  (N^{\rm trap})^2 \sqrt{k_B T} .
  \end{split}
\end{equation}
The first term decreases with increasing $T$, in sharp contrast to the homogeneous case. In a trap, the molecular cloud expands when the temperature increases such that densities and the total contacts decrease for a fixed $N$. Similarly, the second term increases slower than the result in homogenous systems with increasing $T$. Alternatively, we could consider the density at the center of the trap, the decay rate of which linearly depends on $T$ again (Supplementary Materials). 

Though we have used the high temperature regime as an example to explain Eqs. (\ref{decay1}, \ref{decay2}, \ref{nk}, \ref{dcin}), we need to emphasize that these universal relations are powerful tools at any temperatures. In particular, at lower temperatures, contacts are no longer proportional to $N^2$, directly reflecting the critical roles of many-body correlations in determining the reaction rate. For instance, below the superfluid transition temperature, contacts may be directly related to superfluid order parameters \cite{Zhou1, Ueda1}. Universal relations constructed here thus offer us a unique means to explore the interplay between the chemical reaction and symmetry breaking in quantum many-body systems, and more broadly, universality in non-hermitian systems. We hope that our work will stimulate more studies of contacts and universal relations to bridge quantum chemistry, AMO physics, and condensed matter physics.

We thank Jun Ye and Ana Maria Rey for helpful discussions. This work is supported by NSF PHY 1806796 and HKRGC through HKUST3/CRF/13G.

\end{document}

\title{Supplementary Material for ``Universal relations for ultracold reactive molecules"}
\author{Mingyuan He}
%\thanks{They contribute equally to this work.}
\affiliation{Department of Physics and Astronomy, Purdue University, West Lafayette, IN, 47907, USA}
\affiliation{Shenzhen JL Computational Science and Applied Research Institute, Shenzhen, 518000, China}
\affiliation{Department of Physics, The Hong Kong University of Science and Technology, Clear Water Bay, Kowloon, Hong Kong, China}
\author{Chenwei Lv}
%\thanks{They contribute equally to this work.}
\affiliation{Department of Physics and Astronomy, Purdue University, West Lafayette, IN, 47907, USA}
\thanks{They contribute equally to this work.}
\author{Hai-Qing Lin}
\affiliation{Beijing Computational Science Research Center, Beijing, 100193, China}
\author{Qi Zhou}
%\email{zhou753@purdue.edu}
\affiliation{Department of Physics and Astronomy, Purdue University, West Lafayette, IN, 47907, USA}
\affiliation{Purdue Quantum Science and Engineering Institute, Purdue University, West Lafayette, IN, 47907, USA}
\date{\today}
\maketitle

\vspace{0.2in}

\vspace{0.15in}

{\bf The Lindblad equation}

We consider a Lindblad master equation, 
\begin{equation}
\label{eq:master_equation}
\hbar\frac{d\rho}{dt}=-i[H,\rho]+\mathcal{D}[\rho],
\end{equation}
where $H$ is the Hamiltonian that describes the unitary part of the time evolution and the dissipator $\mathcal{D}$ describes the loss due to inelastic collisions,
\begin{eqnarray}
\mathcal{D}[\rho]&=&-\int d^3x_1d^3x_2 \frac1{2}\Gamma(|{\bf x_1-x_2}|)\left(2\Psi({\bf x_2})\Psi({\bf x_1})\rho\Psi^\dagger({\bf x_1})\Psi^\dagger({\bf x_2})-\{\Psi^\dagger({\bf x_1})\Psi^\dagger({\bf x_2})\Psi({\bf x_2})\Psi({\bf x_1}),\rho\}\right), \label{eq:dissipator}
\end{eqnarray}
$\Psi(\bf{x})$ is the fermionic field operator satisfying $\{\Psi({\bf x}),\Psi^\dagger({\bf x'})\}=\delta^{(3)}({\bf x}-{\bf x'})$. $( 1/2)\Gamma(|{\bf x}_1-{\bf x}_2|)$ describes a finite range dissipation. The loss rate of the total particle number, $dN/ dt=\int d^3x(d/dt){\rm Tr}( n({\bf x})\rho)$, $n({\bf x})=\Psi^\dagger({\bf x})\Psi({\bf x})$, is written as 
\begin{equation}\label{eq:nd_me}
  \begin{split}
    \frac{dN}{dt}=&-\frac{1}{\hbar}{\rm Tr}\Big(\int d^3x\Psi^\dagger({\bf x})\Psi({\bf x})\int d^3x_1d^3x_2 \frac{1}{2} \Gamma(|{\bf x}_1-{\bf x}_2|)\left[2\Psi({\bf x}_2)\Psi({\bf x}_1)\rho\Psi^\dagger({\bf x}_1)\Psi^\dagger({\bf x}_2)\right. \\
      & \qquad \qquad\qquad \qquad\qquad \qquad\qquad \qquad\qquad \qquad\qquad\quad \left.-\{\Psi^\dagger({\bf x}_1)\Psi^\dagger({\bf x}_2)\Psi({\bf x}_2)\Psi({\bf x}_1),\rho\}\right]\Big)\\
    =&-\frac{1}{\hbar}\int d^3x_1d^3x_2 d^3x\Gamma(|{\bf x}_1-{\bf x}_2|){\rm Tr}\left(\rho[\Psi^{\dagger}({\bf x}_1)\Psi^{\dagger}({\bf x}_2),\Psi^\dagger({\bf x})\Psi({\bf x})]\Psi({\bf x}_2)\Psi({\bf x}_1)\right)\\
    =&\frac2{\hbar}\int d^3xd^3x'  \Gamma(|{\bf x'-x}|)\big\langle \Psi^{\dagger}({\bf x})\Psi^{\dagger}({\bf x'})\Psi({\bf x'})\Psi({\bf x})\big\rangle.
  \end{split}
\end{equation}
This equation is valid for any finite range dissipator. In the approximation of zero-range dissipators, $\Gamma= g \delta^{(3)} ({\bf x}-{\bf x'})$, it reduces to \cite{Eric1,Rempe1,Cirac1}
\begin{equation}\label{eq:nd_contact}
  \frac{dN}{dt}=\frac2{\hbar}g\int d^3x\big\langle \Psi^{\dagger}({\bf x})\Psi^{\dagger}({\bf x})\Psi({\bf x})\Psi({\bf x})\big\rangle.
\end{equation}
Reference \cite{Eric1} considered two-component fermions and obtained $d\langle N_1\rangle/dt=d\langle N_2\rangle/dt=-[\hbar/(2\pi m)] {\rm Im}(1/a)C$, where $N_1$ ($N_2$) is the number of spin-up (spin-down) fermions, $a$ is the s-wave scattering length,  and $C$ is the s-wave contact. 

We emphasize that Eq. (\ref{eq:nd_me}) is equivalent to results derived from the Hamiltonian with a complex interaction, as shown in Eq. (\ref{eq:nd_cp_fq}) in the next section, provided that we identify $U_I$ and $\Gamma$, i.e., $U_I(|{\bf x'-x}|)=\Gamma(|{\bf x'-x}|)$. Thus, universal relations derived from the Lindblad equation are the same as those shown in the main text, since the probability of having more than two particles within a distance smaller than $r_0$ is negligible in dilute systems satisfying $r_0\ll k_F^{-1}$.  \\

{\bf Decay rate}

In the presence of complex short-range interactions, a many-body wavefunction satisfies
\begin{equation}
i\hbar \partial_t \Psi({\bf r}_1, {\bf r}_2, ..., {\bf r}_N)=\Big[\sum_{i=1}^{N} [-\frac{\hbar^2}{2M}\nabla_i^2+V_{\rm ext}({\bf r}_i)]+\sum_{i>j} U({\bf r}_{ij}) \Big] \Psi({\bf r}_1, {\bf r}_2, ..., {\bf r}_N).
\end{equation}
For any finite size system, net current vanishes at the boundary. We obtain
\begin{equation}
\label{decay}
\partial_t N=\frac{4}{\hbar} \sum\nolimits_{i>j}  \int d{\bf R}_{ij} d{\bf r}_{ij} U_I({\bf r}_{ij})|\Psi({\bf R}_{ij},{\bf r}_{ij} )|^2,
\end{equation}
which is equivalent to a second quantization form using fermionic operators, 
\begin{equation}\label{eq:nd_cp_fq}
  \begin{split}
\partial_t N=\frac2{\hbar}\int d^3xd^3x'  U_I(|{\bf x'-x}|)\big\langle \Psi^{\dagger}({\bf x})\Psi^{\dagger}({\bf x'})\Psi({\bf x'})\Psi({\bf x})\big\rangle.
  \end{split}
\end{equation}

Using Eq. (3) in the main text, and $\epsilon \psi_m({\bf r}_{ij};\epsilon) = [-({\hbar^2}/{M}) \nabla_{r_{ij}}^2+ U({\bf r}_{ij})] \psi_m({\bf r}_{ij};\epsilon)$, we obtain
\begin{equation}
\label{CEM1}
\begin{split}
   & 2i\sum\nolimits_{j > i} {\int {d{{\bf R}_{ij}}\int_0^{{r_0}} {d{{\bf r}_{ij}}|\Psi \left( {{{\bf R}_{ij}},{{\bf r}_{ij}}} \right)|^2 {U_I}\left( {{{\bf r}_{ij}}} \right)} } } \\
   &- \sum\nolimits_{j > i} {\int {d{{\bf R}_{ij}}\int_0^{{r_0}} {d{{\bf r}_{ij}}{\Psi ^*}\left( {{{\bf R}_{ij}},{{\bf r}_{ij}}} \right)\sum\nolimits_{m,\epsilon } {\epsilon {G_m}\left( {{{\bf R}_{ij}};E - \epsilon } \right){\psi _m}({{\bf r}_{ij}};\epsilon )} } } } \\
   &+ \sum\nolimits_{j > i} {\int {d{{\bf R}_{ij}}\int_0^{{r_0}} {d{{\bf r}_{ij}}{\Psi }\left( {{{\bf R}_{ij}},{{\bf r}_{ij}}} \right)\sum\nolimits_{m,\epsilon} {\epsilon^* {G_m^*}\left( {{{\bf R}_{ij}};E - \epsilon } \right){\psi _m^*}({{\bf r}_{ij}};\epsilon )} } } }  \\
 =& \frac{{{\hbar ^2}}}{M}\sum\nolimits_{j > i} {\int {d{{\bf R}_{ij}}\int_0^{{r_0}} {d{{\bf r}_{ij}}\left[ {{\Psi ^*}\left( {{{\bf R}_{ij}},{{\bf r}_{ij}}} \right)\nabla _{{r_{ij}}}^2\Psi \left( {{{\bf R}_{ij}},{{\bf r}_{ij}}} \right) - \Psi \left( {{{\bf R}_{ij}},{{\bf r}_{ij}}} \right)\nabla _{{r_{ij}}}^2{\Psi ^*}\left( {{{\bf R}_{ij}},{{\bf r}_{ij}}} \right)} \right]} } }.
\end{split}
\end{equation}

Note that, for the system with isotropic interactions $U({\bf r})=U(|{\bf r}|)$ and $\psi_{m}({\bf r}_{ij};\epsilon)=\varphi(|{\bf r}_{ij}|;\epsilon)Y_{1m}(\hat{\bf r}_{ij})$, one has \cite{Zhang1}
\begin{eqnarray}
  v_p & = & \frac{r_0^3}{3} \frac{r_0\varsigma-2}{r_0 \varsigma +1}\label{vp},\\
  \frac{1}{r_e}& = & -\frac{1}{r_0}-\frac{r_0^2}{3}\frac{1}{v_p} + \frac{r_0^5}{45} \frac{1}{(v_p)^2} - \int_0^{r_0}  [\varphi^{(0)} (r) ]^2 r^2dr \label{re},
\end{eqnarray}
where $\varsigma=\{\partial \ln [r\varphi(r;\epsilon)]/\partial r\} |_{\epsilon=0}$. Based on Eqs. (4, 5) in the main text and Eq. (\ref{re}), we define
\begin{equation}
C_{1m}^{ss'} = {\left( {4\pi } \right)^2}N\left( {N - 1} \right)\int {d{{\bf R}_{ij}}{{g_{m}^{s*}} }g_{m}^{s'}},\quad g^s_m=\sum\nolimits_\epsilon {q_\epsilon^{2s}G_m({\bf R}_{ij};E-\epsilon)} , \quad {C_{1m}} = C_{1m}^{00},
\end{equation}
and obtain
\begin{equation}
\begin{split}
   & \sum\nolimits_m {\left[ {{\rm{Im}}\left( -v_p^{-1}  \right){C_{1m}} + \frac{1}{2}{\rm{Im}}\left( {r_e^{-1}}  \right)\left( {C_{1m}^{01} + C_{1m}^{10}} \right) - i \kappa_3 \left( {C_{1m}^{01} - C_{1m}^{10}} \right) + {\rm{O}}\left( {{{(E - {E^*})}^2}} \right)} \right]}  \\
 =&\frac{\left(4\pi\right)^2 2M}{\hbar^2} \sum\nolimits_{i>j}\int {d{{\bf R}_{ij}}\int_0^{{r_0}} {d{{\bf r}_{ij}}{{\left| \sum\nolimits_{m,\epsilon}{G_{m}}\left( {{{\bf R}_{ij}};E - \epsilon} \right)\psi_m({{\bf r}_{ij}};{\epsilon})  \right|}^2}U_I\left( {{{\bf r}_{ij}}} \right)} }  ,
 \end{split}
\end{equation}
which leads to Eqs. (10, 11, 12) in the main text,
\begin{eqnarray}
\kappa_1&=&{\rm Im}\left(v_p^{-1} \right)=-\frac{M}{{{\hbar ^2}}}\int_0^{\infty} {d{ r}r^2{{\left| \varphi^{(0)}({r})\right|}^2}U_I({r})}, \\
\kappa_2&=&{\rm{Im}}\left( -{r_e^{-1}}/2 \right) =-\frac{{M}}{{{\hbar ^2}}}{\mathop{\rm Re}\nolimits} \left( {\int_0^\infty  {d{r} r^2\varphi ^{(0)*}({r})\varphi ^{(1)}(r)U_I({r})} } \right), \\
\kappa_3&=&-\int_0^{{r_0}} {\left\{ {{{\left[ {{\rm{Im}}\varphi^{(0)}\left( r \right)} \right]}^2} - {{\left[ {{\mathop{\rm Im}\nolimits} {\tilde \varphi^{(0)}}\left( r \right)} \right]}^2}} \right\}r^2dr}
= \frac{{M}}{{{\hbar ^2}}}{\mathop{\rm Im}\nolimits} \left( {\int_0^\infty  {d{r} r^2\varphi^{(0)*}({r})\varphi^{(1)}(r)U_I({r})} } \right),
\end{eqnarray}
where ${\tilde \varphi^{(0)}}\left( r \right)$ is a wave function obtained from extending the actual wave function ${\varphi^{(0)}}\left( r \right)$ outside the potential ($r>r_0$) into the regime 
$r<r_0$. We obtain Eq. (13) in the main text,
\begin{equation}
\label{decayN}
\partial_t N= -\frac{\hbar}{8\pi^2 M}\Big[\text{Im} ({v^{-1}_p})C_1 - \frac{1}{2}\text{Im} (r_{e}^{-1}) C_2+\kappa_3 C_3\Big],
\end{equation}
where $C_1=\sum\nolimits_m C_{1m}$, $C_2=2\sum\nolimits_m {\rm Re}(C_{1m}^{01})$ and $C_3=2\sum\nolimits_m {\rm Im}(C_{1m}^{01})$. The above equation leads to Eqs. (7, 8, 9) in the main text by considering $G_m({\bf R}_{ij};E-\epsilon)=G({\bf R}_{ij};E-\epsilon)$.\\

{\bf Momentum distribution}

Similar to systems with real interactions \cite{He}, using  $n\left( {\bf k} \right) = \sum\nolimits_{i = 1}^N {\int {\prod\nolimits_{j \ne i} {d{{\bf r}_j}} |\int {d{{\bf r}_i}\Psi \left( {{{\bf r}_1},{{\bf r}_2}, ..., {{\bf r}_N}} \right){e^{ - i{\bf k} \cdot {{\bf r}_i}}}} {|^2}} } $, We obtain
\begin{equation}
n({\bf k}) \stackrel{k_F\ll |{\bf k}|\ll r_0^{-1}}{\xrightarrow{\hspace*{1.5cm}} } \frac{C_1}{3|{\bf k}|^{2}} \sum_{m}  |Y_{1m}(\hat {\bf k})|^2,
\end{equation}
which leads to 
\begin{equation}
n(|{\bf k}|)=\int d{\bf \Omega} n({\bf k})= \frac{C_1}{|{\bf k}|^{2}}. 
\end{equation}

{\bf Density correlations}

The density correlation function $S({\bf r})=\int d{\bf R} \langle n({\bf R}+{\bf r}/2)n({\bf R}-{\bf r}/2)\rangle$ can be rewritten as 
\begin{equation}\label{eq:density_correlation1}
  S({\bf r})={N(N-1)} \int\left(\prod\nolimits_{k \neq i, j} d{\bf r}_{k}\right)
  {\left|\Psi\left(\mathbf{r}_{1}, ..., \mathbf{r}_{i}=\mathbf{R}+\frac{\mathbf{r}}{2}, ..., \mathbf{r}_{j}=\mathbf{R}-\frac{\mathbf{r}}{2}, ..., \mathbf{r}_{N}\right)\right|^{2}}.
\end{equation} 
In the regime,  $r\ll k_F^{-1}$, $S({\bf r})$ is written as 
\begin{equation} 
\label{eq:density_correlation3}
S(\mathbf{r})={N(N-1)}\int d{\bf R}_{ij}\sum_{m}  |Y_{1m}(\hat{\bf {r}})|^2\left[ |\varphi^{(0)}_{m}(r)|^2 |g^{(0)}_{m}|^2+\varphi^{(0)}_{m}(r)\varphi^{(1)*}_{m}(r)g^{(1)*}_{m}g^{(0)}_{m}+\varphi^{(1)}_{m}(r)\varphi^{(0)*}_{m}(r)g^{(0)*}_{m}g^{(1)}_{m}\right].
\end{equation} 
where $r=|{\bf r}|$. The integral over a shell allows us to obtain 
\begin{equation}
    P(x, D)=\int_{x}^{x+D} d{\bf r} S({\bf r})=\frac{1}{(4\pi)^2}\sum_{m}\int_{x}^{x+D}r^2dr\left[
    \varphi^{(0)}_{m}(r)\varphi^{(0)*}_{m}(r)C_{1m}+\varphi^{(0)}_{m}(r)\varphi^{(1)*}_{m}(r)C_{1m}^{10}+\varphi^{(1)}_{m}(r)\varphi^{(0)*}_{m}(r)C_{1m}^{01}\right]. \label{eq:integrated_density_correlation}
\end{equation}
Using Eqs. (4, 5) in the main text, we obtain
\begin{equation} \label{eq:density_correlation_r}
  \begin{split}
    \left.{\frac{\partial P(x, D)}{\partial D}}\right|_{D\to0}
   =&\frac{1}{16\pi^2} \left\{ C_1\frac1{x^2}+C_2\frac{1}{2}+\left[-2{\rm Re}(\frac1{v_p})C_1+{\rm Re}(\frac1{r_e})C_2-{\rm Im}(\frac1{r_e})C_3\right]\frac{x}{3} \right.\\
    &\left. +\left[-\frac{2}{3}{\rm Re}(\frac1{v_p})C_2-{\rm Im}(\frac1{v_p})C_3\right]\frac{x^3}{5}+\left[\frac{1}{|v_p|^2}C_1-{\rm Re}(\frac1{v_p^*r_e})C_2+{\rm Im}(\frac1{v_p^*r_e})C_3\right]\frac{x^4}{9}-\frac{C_2}{90|v_p|^2}x^6\right\},
  \end{split}
\end{equation}
where the first line recovers Eq. (15) in the main text, and the second line includes higher order terms. When $v_p$ and $r_e$ are known, the first line readily allows experimentalists to obtain $C_{1,2,3}$ by fitting the experimental data. When $v_p$ and $r_e$ are unknown, the second line is required to obtain $C_{\nu}$, $v_p$ and $r_e$. \\

{\bf Finite temperature results}

\begin{figure}
  \centering
  \includegraphics[width=0.7\textwidth]{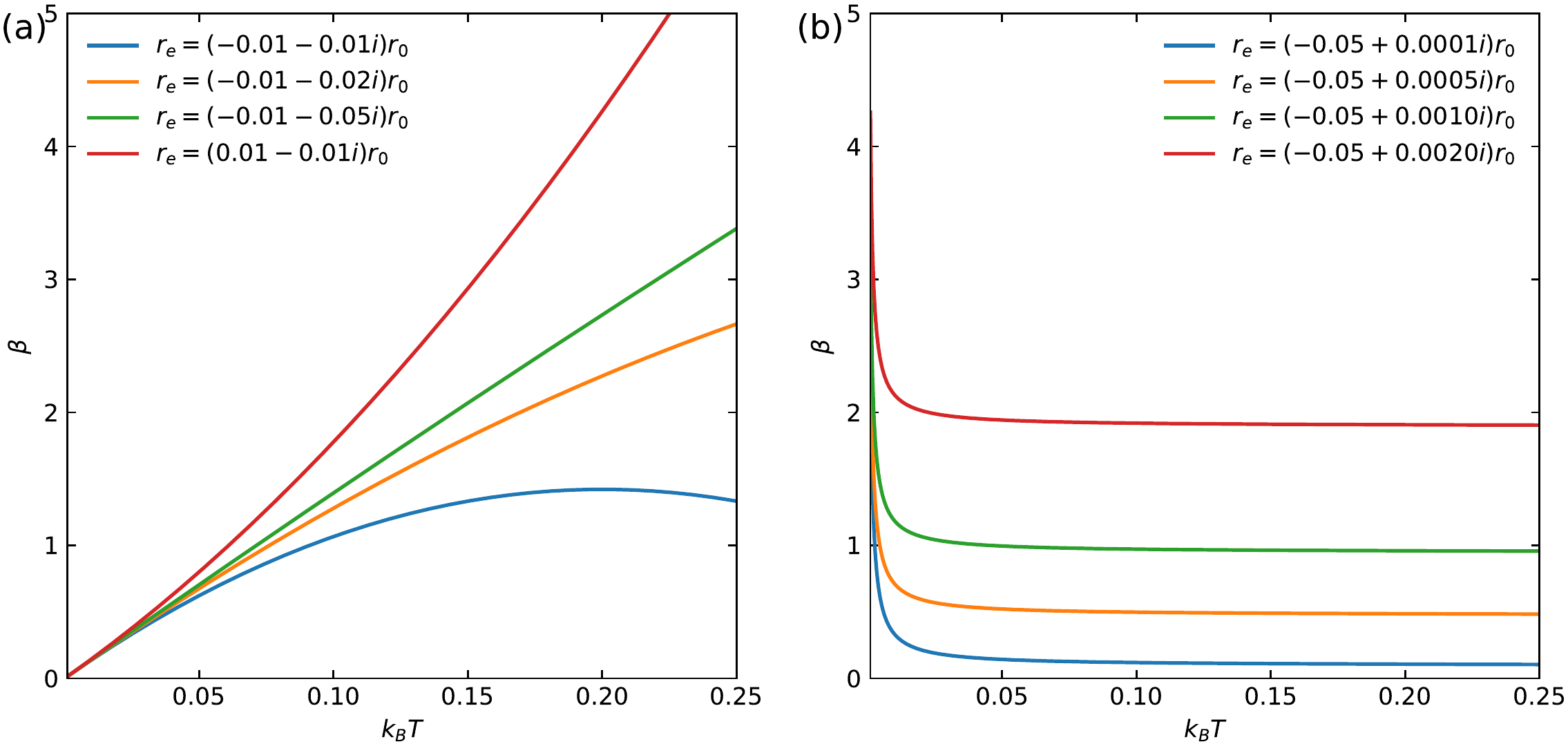}
  \caption{Decay rate $\beta$ (in unit of $[{r_0^3}/({Vh})][\hbar^2/(Mr_0^2)]$) in a homogenous system as a function of $k_BT$ (in unit of $\hbar^2/(Mr_0^2)$) at different effective ranges. (a) For the limit $v_p \to 0$, $v_p$ is fixed at $(0.01-0.01i)r_0^3$. (b) For the limit $v_p \to \infty$, $v_p$ is fixed at $(0.01-2000i)r_0^3$.    }\label{Fig1s}
\end{figure}
  
 {\it Homogenous systems}
 
We define the decay rate, $\beta$,  using
\begin{equation}
  \partial_t N=-\beta N^2. 
\end{equation}
For a two-body scattering state, we obtain
\begin{eqnarray}
\left. C_1\right|_{v_p\to 0} & = & 12(4\pi)^3 q_\epsilon^2 |v_p|^2 \left[1+2{\rm Re} (v_p/r_e) q_\epsilon^2\right]/V,\\
\left. C_2\right|_{v_p\to 0} & = & 24(4\pi)^3 q_\epsilon^4 |v_p|^2 /V,\\
\left. C_1\right|_{v_p\to \infty} & = & 12(4\pi)^3 q_\epsilon^{-2} |r_e|^2 /V, \\
\left. C_2\right|_{v_p\to \infty} & = & 24(4\pi)^3 |r_e|^2 \left[1+2{\rm Re} (r_e/v_p) q_\epsilon^{-2} \right] /V.
\end{eqnarray}
Thermal averaged results are written as
\begin{eqnarray}
\left. \langle C_1\rangle_T\right|_{v_p\to 0} & = & \frac{9}{2}(4\pi)^4 N n |v_p|^2\lambda_T^{-2} \left[1+10\pi {\rm Re} (v_p/r_e)\lambda_T^{-2} \right],\\
\left. \langle C_2\rangle_T\right|_{v_p\to 0} & = & \frac{45}{4}(4\pi)^5  N n |v_p|^2 \lambda_T^{-4},\\
\left. \langle C_1\rangle_T\right|_{v_p\to \infty} & = & 24(4\pi)^2 N n |r_e|^2 \lambda_T^{2}, \\
\left. \langle C_2\rangle_T\right|_{v_p\to \infty} & = & 12(4\pi)^3 N n |r_e|^2 \left[1+\frac{2}{\pi}{\rm Re} (r_e/v_p) \lambda_T^{2} \right] .
\end{eqnarray}
When the bound state is not occupied, $\langle C_3 \rangle_T=0$, we obtain
\begin{eqnarray}
\left. \beta \right|_{v_p\to 0}(T) & = & -\frac{r_0^3}{Vh}  144\pi^2\left[{\rm Im}\left(\frac{v_p}{r_0^3} \right) k_B T+\frac{5}{2}\left|\frac{v_p}{r_0^3}\right|^2{\rm Im}\left(\frac{v_p}{v_p^*}\frac{r_0}{r_e}\right) \frac{Mr_0^2} {\hbar^2}k_B^2 T^2\right],\\
\left. \beta \right|_{v_p\to \infty}(T) & = &  -\frac{r_0^3}{Vh} \frac{\hbar^2}{Mr_0^2} 96\pi^2\left|\frac{r_e}{r_0}\right|^2\left[{\rm Im}\left(\frac{r_0}{r_e} \right)+2 {\rm Im}\left(-\frac{r_e}{r_e^*}\frac{r_0^3}{v_p}\right)\frac{\hbar^2}{Mr_0^2}\frac{1}{k_B T}\right].
\end{eqnarray}
In figure \ref{Fig1s}, we plot the decay rate as a function of $T$ for both small and large scattering volumes.\\

{\it Harmonic traps} 

In the high temperature regime, the density can be well approximated by $n=\exp[{\mu/(k_BT)}]\lambda_T^{-3}$. Applying the local density approximation in a harmonic trap, $\mu$ is replaced by the local chemical potential $\mu({\bf r})=\mu(0)-V_{\rm ext}({\bf r})$, $V_{\rm ext}({\bf r})=(1/2)M\omega^2(x_1^2+x_2^2+ x_3^2) $, $\omega$ is the harmonic frequency, and $n({\bf r})=\exp[{\mu({\bf r})/(k_B T)}]\lambda_T^{-3}$ where $\mu(0)$ is the chemical potential at the center of the trap and is fixed by the total particle number, $\int d^3 r n({\bf r})=N^{\rm trap}$.  Within this framework, the density at the center of the trap is determined by $N^{\rm trap}$ and $T$, $n(0)=  [({2\pi k_B T})/({M \omega^2})]^{-3/2} N^{\rm trap}$. The total contacts are obtained from integrating the local contacts in the trap,
\begin{equation}
\langle C_\nu^{\rm trap}\rangle_T= {\cal C}_\nu(0) \int e^{-2V_{\rm ext} ({\bf r})/(k_B T)}d{\bf r} = \left(\frac{\pi k_B T}{M\omega^2}\right)^{3/2} {\cal C}_\nu(0).
\end{equation}
where ${\cal C}_\nu=\langle C_\nu/V\rangle_T $ is the contact density of $\nu$th contact at the center of the trap.

\begin{figure}
  \centering
  \includegraphics[width=0.7\textwidth]{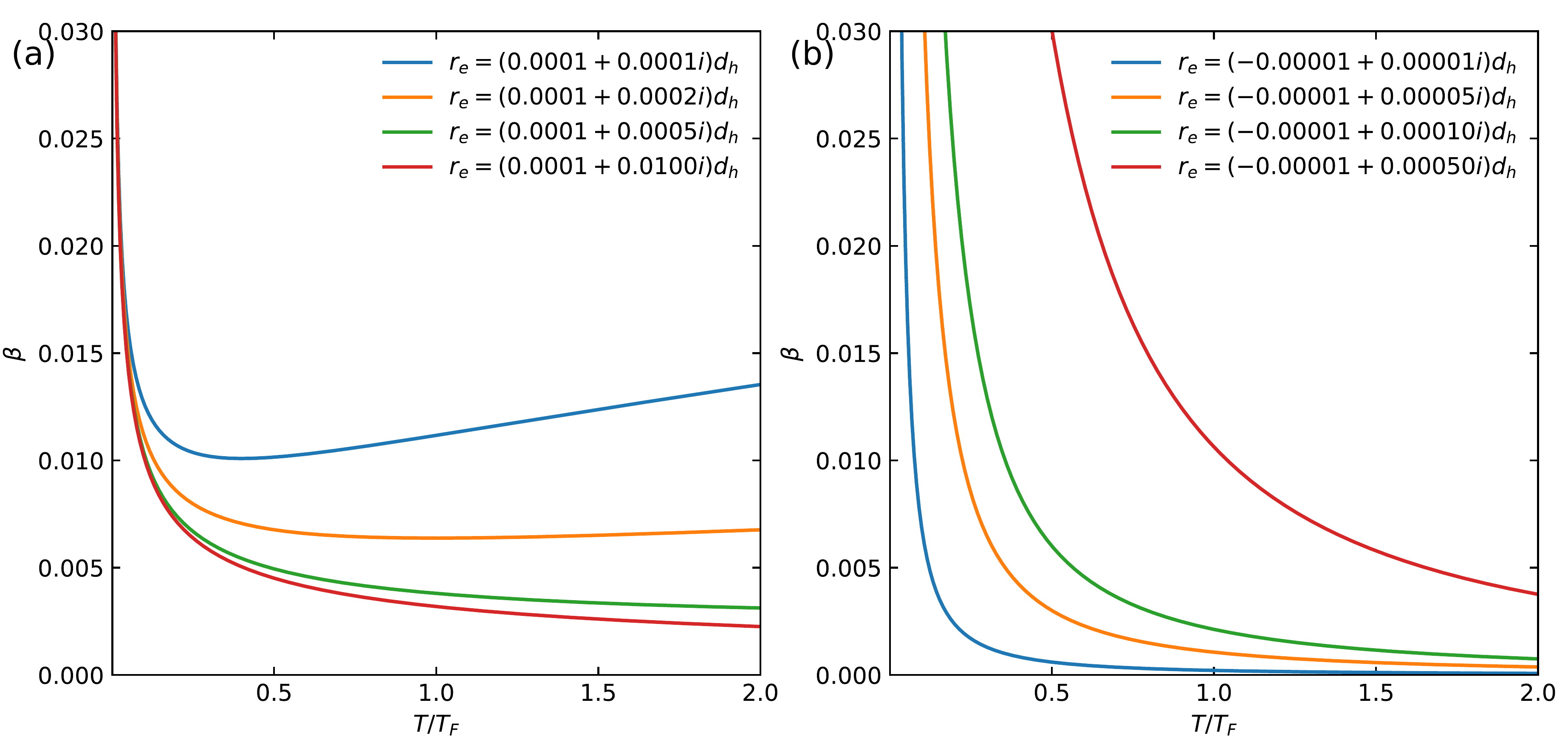}
  \caption{Decay rate $\beta$ (in unit of $(\hbar{\omega})/h$) in a harmonic trap as a function of $T/T_F$ (in unit of $k_F^{-2}d_h^{-2}$) at the limit $v_p \to 0$ for different effective ranges, (a) For the limit $v_p \to 0$, $v_p$ is fixed at $(0.0001-0.0001i)d_h^3$. (b) For the limit $v_p \to \infty$, $v_p$ is fixed at $(0.01-20i)d_h^3$.  }\label{Fig2s}
\end{figure}

\begin{figure}
  \centering
  \includegraphics[width=0.7\textwidth]{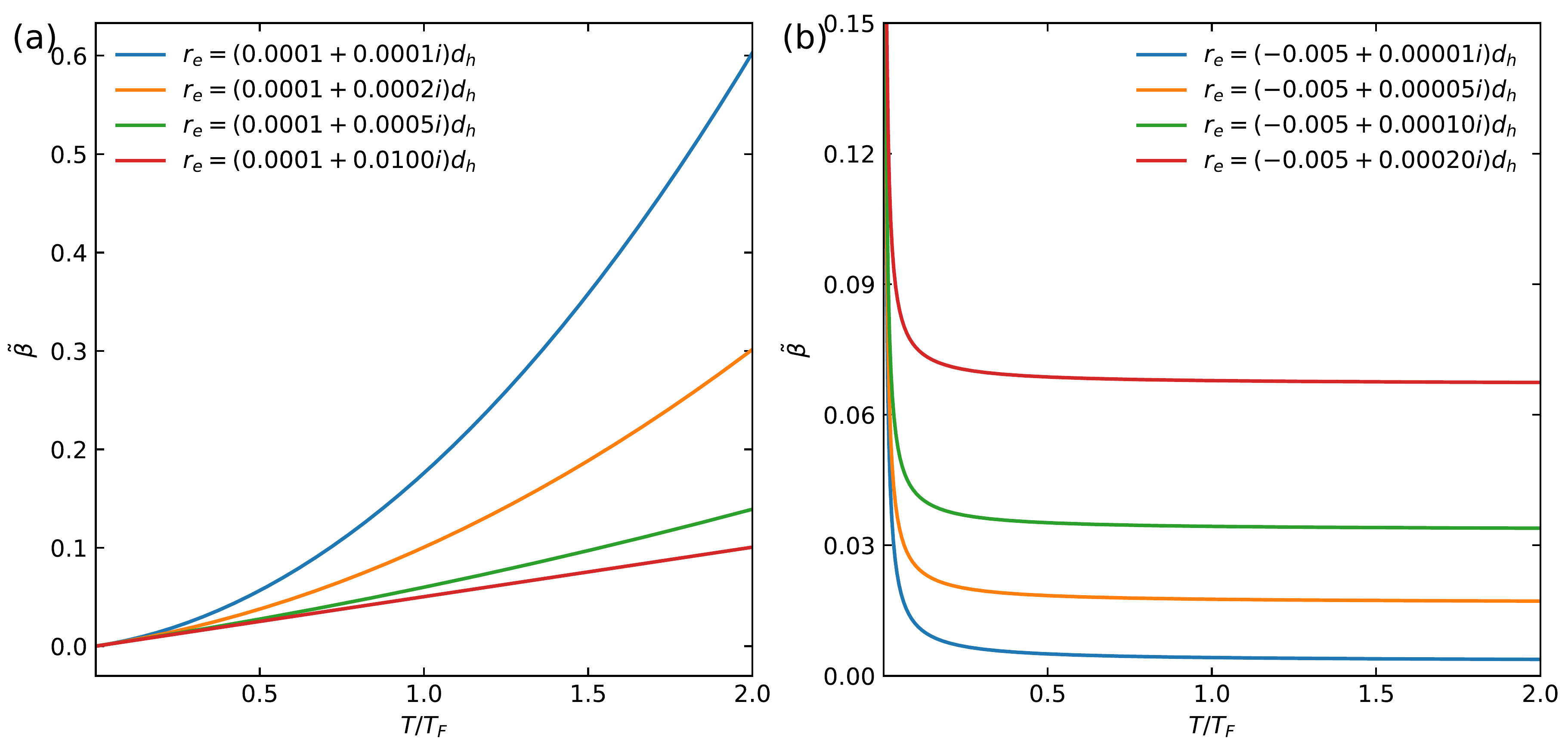}
  \caption{ Decay rate $\tilde{\beta}$ (in unit of $(d_h^3(\hbar{\omega}/h)$) for $n(0)$ at the center of a harmonic trap as a function of $T/T_F$ (in unit of $k_F^{-2}d_h^{-2}$) at the limit $v_p \to 0$ for different effective ranges, (a) For the limit $v_p \to 0$, $v_p$ is fixed at $(0.0001-0.0001i)d_h^3$. (b) For the limit $v_p \to \infty$, $v_p$ is fixed at $(0.01-20i)d_h^3$.  }\label{Fig3s}
\end{figure}

In the weak interaction limit and without the bound state, one has $\langle C_3^{\rm trap}\rangle_T =0$, and
\begin{eqnarray}
\langle C_1^{\rm trap}\rangle_T & = & 72 \pi^{3/2} |v_p|^2 d_h^{-5} (N^{\rm trap})^2\left( \frac{1}{k_F^2 d_h^2}\frac{T_F}{T}\right)^{1/2}\left[1+5{\rm Re} \left(\frac{v_p}{d_h^3}\frac{d_h}{r_e}\right)   k_F^2 d_h^2\frac{T}{T_F} \right], \\
\langle C_2^{\rm trap}\rangle_T & = & 360 \pi^{3/2} |v_p|^2 d_h^{-7} (N^{\rm trap})^2\left( k_F^2 d_h^2\frac{T}{T_F}\right)^{1/2},
\end{eqnarray}
where $k_F=(2M\omega/\hbar)^{1/2}(3N^{\rm trap})^{1/6}$, $T_F=(\hbar \omega/k_B)(3N^{\rm trap})^{1/3}$, and $d_h=[\hbar/(M\omega)]^{1/2}$. The decay rate is then
\begin{equation}
\left. \beta \right|_{v_p\to 0}(T)  = 18\sqrt{\pi}\frac{\hbar\omega}{h} \left|\frac{v_p}{d_h^3}\right|^2\left[{\rm Im}\left(\frac{d_h^3}{v_p} \right)\frac{1}{k_F d_h} \sqrt{\frac{T_F}{T}}-\frac{5}{2} {\rm Im}\left(\frac{v_p}{v_p^*}\frac{d_h}{r_e}\right) {k_F d_h} \sqrt{\frac{T}{T_F}} \right].
\end{equation}

In the strong interaction limit, one has $\langle C_3^{\rm trap}\rangle_T =0$, and
\begin{eqnarray}
  \langle C_1^{\rm trap}\rangle_T & = & 96 \pi^{3/2} |r_e|^2 d_h^{-1}(N^{\rm trap})^2\left( \frac{1}{k_F^2 d_h^2}\frac{T_F}{T}\right)^{5/2}, \\
  \langle C_2^{\rm trap}\rangle_T & = & 96 \pi^{3/2} |r_e|^2 d_h^{-3}(N^{\rm trap})^2 \left( \frac{1}{k_F^2 d_h^2}\frac{T_F}{T}\right)^{3/2}\left[1+4 {\rm Re} \left(\frac{d_h^3}{v_p}\frac{r_e}{d_h}\right)\left( \frac{1}{k_F^2 d_h^2}\frac{T_F}{T}\right)\right].
\end{eqnarray}
The decay rate is then
\begin{equation}
  \left. \beta \right|_{v_p\to \infty}(T)  =  -12\sqrt{\pi}\frac{\hbar{\omega}}{h}\left|\frac{r_e}{d_h}\right|^2\left[{\rm Im}\left(\frac{d_h}{r_e} \right) \frac{1}{k_F^3d_h^3} \left(\frac{T_F}{T}\right)^{3/2} +2 {\rm Im}\left(-\frac{r_e}{r_e^*}\frac{d_h^3}{v_p}\right) \frac{1}{k_F^5d_h^5} \left(\frac{T_F}{T}\right)^{5/2}\right].
\end{equation}
Figure \ref{Fig2s} shows the decay rate as a function of $T$ in a trap. If the decay rate at the center of the trap $\tilde{\beta}$, which satisfies
\begin{equation}\label{decay_ctrap}
  \partial_tn(0)=-\tilde{\beta}n^2(0),
\end{equation}
 is considered, one has
 \begin{eqnarray}
  \left. \tilde{\beta} \right|_{v_p\to 0}(T)  &=& 36\sqrt{2}\pi^2d_h^3\frac{\hbar\omega}{h} \left|\frac{v_p}{d_h^3}\right|^2\left[{\rm Im}\left(\frac{d_h^3}{v_p} \right)k_F^2d_h^2\frac{T}{T_F} -\frac{5}{2} {\rm Im}\left(\frac{v_p}{v_p^*}\frac{d_h}{r_e}\right)k_F^4d_h^4\frac{T^2}{T_F^2}\right], \label{Trap_decay_n0}\\
  \left. \tilde{\beta}\right|_{v_p\to \infty}(T) & =  &-24\sqrt{2}\pi^2d_h^3\frac{\hbar{\omega}}{h}\left|\frac{r_e}{d_h}\right|^2\left[{\rm Im}\left(\frac{d_h}{r_e} \right)+2 {\rm Im}\left(-\frac{r_e}{r_e^*}\frac{d_h^3}{v_p}\right)\frac{1}{k_F^2d_h^2}\frac{T_F}{T} \right].
\end{eqnarray}
The decay rate as a function of $T$ at the center of the trap is shown in figure \ref{Fig3s}. \\

{\bf Measure $\kappa_{1,2,3}$ in a two-body system}

In a two-body system, $\epsilon$ is a good quantum number such that the three contacts are not independent. In fact, $C_2=2C_1{\rm Re}(q_\epsilon ^2)$ and $C_3=2C_1{\rm Im}(q_\epsilon^2)$. Therefore, the decay rate is determined by \begin{equation}
\partial_t N= - \frac{\hbar C_1}{8\pi ^2 M}[\kappa_1 + 2{\rm Re}(q_\epsilon^2)\kappa_2 + 2{\rm Im}(q_\epsilon ^2)\kappa_3  ]. \label{decay1s}
\end{equation}
$C_1$ can be measured from the momentum distribution using $n(|{\bf k}|) \stackrel{|q_\epsilon|\ll |{\bf k}|\ll r_0^{-1}}{\xrightarrow{\hspace*{1.5cm}} } {C_1}/{|{\bf k}|^2}$. In the above equation, except $\kappa_{1,2,3}$, all other quantities are measurable. Therefore, $\kappa_{1,2,3}$ can be extracted from the experimental data. To be more explicit, in a two-body system, $\partial_t N$ is determined by ${\rm Im}(\epsilon)$. Equation (\ref{decay1s}) is rewritten as $\partial_t N =-[(\hbar C_1)/(8\pi^2M)] [\kappa_1 +2 {\rm Re}(q_\epsilon^2) \kappa_2]/[1+C_1 \kappa_3/(16\pi^2) ]$. Thus, from the decay rate, the momentum distribution and the real part of the energy that can be obtained from a spectroscopy measurement, all three microscopic parameters, $\kappa_{1,2,3}$ can be accessed in experiments.\\ 

{\bf The decay rate of the density at the center of the trap}

If we consider the density at the center of the trap, $n(0)=[(2\pi k_B T)/(M\omega^2)]^{-3/2}N^{\rm trap}$, in the weakly interacting regime, based on Eq. (\ref{Trap_decay_n0}), the differential equation satisfied by $n(0)$ is written as 
\begin{equation}
\begin{split}
\partial_t n(0)=&\frac{36\sqrt{2}\pi^2}{h}{\rm Im}(v_p)n^2(0) k_BT +\frac{90\sqrt{2}\pi^2}{h}  {\rm Im} (\frac{v_p}{v_p^*} r_e^{-1}) \frac{M |v_p|^2}{\hbar^2}  n^2(0) k_B^2 T^2. 
\end{split}
\end{equation}
We see that the first term becomes linearly dependent on $T$. This is what has been observed in the recent experiment, which has defined an average density proportional to $n(0)$ \cite{Ye1}. Taking into account the second term, there will be a deviation from the linear dependence on $T$.